# MODELING UPCONVERSION OF ERBIUM DOPED MICROCRYSTALS BASED ON EXPERIMENTALLY DETERMINED EINSTEIN COEFFICIENTS

S. Fischer[1], H. Steinkemper[1], P. Löper[1], M. Hermle[1], J. C. Goldschmidt[1]

[1]Fraunhofer Institute for Solar Energy Systems, Heidenhofstr. 2, 79110 Freiburg, Germany

Phone: +49 (0) 761 4588 5955 Fax: + 49 (0) 4588 9250 Email: stefan.fischer@ise.fraunhofer.de

## Abstract

Upconversion of infrared photons is a promising possibility to enhance solar cell efficiency by producing electricity from otherwise unused sub-band-gap photons. We present a rate equation model, and the relevant processes, in order to describe upconversion of near-infrared photons. The model considers stimulated and spontaneous processes, multi-phonon relaxation and energy transfer between neighboring ions. The input parameters for the model are experimentally determined for the material system $\beta$-NaEr$_{0.2}$Y$_{0.8}$F$_4$. The determination of the transition probabilities, also known as the Einstein coefficients, is in the focus of the parameterization. The influence of multi-phonon relaxation and energy transfer on the upconversion are evaluated and discussed in detail. Since upconversion is a non-linear process, the irradiance dependence of the simulations is investigated and compared to experimental data of quantum efficiency measurements. The results are very promising and indicate that upconversion is physically reasonably described by the rate equations. Therefore, the presented model will be the basis for further simulations concerning various applications of upconversion, such as in combination with plasmon resonances in metal nanoparticles.

PACS numbers: 42.65.Ky, 78.20.Bh, 32.80.-t, 78.55.-m

## I. INTRODUCTION

The properties of rare earth materials are very interesting for many applications, including optical waveguide amplifiers [1,2], lasers [3-6] and biological markers [7,8]. One especially interesting property is their ability to generate high energy photons out of a larger number of low energy photons via multi photon absorption and energy transfer (ET). This generation of higher energy photons is also known as upconversion (UC). UC may have a useful application in photovoltaics (PV). A first approach of a photovoltaic device with upconversion was presented by Gibart et al. [9]. The upconverter was attached on the rear of a GaAs solar cell.

Another opportunity to enhance the efficiency of a photovoltaic device is downshifting or downconversion. The efficiency of the solar cell is enhanced by downshifting because a high energy photon is shifted to a lower energy where the external quantum efficiency of the solar cell is larger. Downconversion (photon cutting) is the opposite process of upconversion. Hence, several lower energy photons are produced from one high energy photon which leads to possible quantum efficiencies far above 100%. Since, a donwconverter needs to be placed on the front of the solar cell several loss mechanisms are introduced to the device and makes the realization of solar cells with downconversion more challenging [10]. The isotropic emission of the luminescent photons, parasitic absorption, concentration quenching and scattering of the incident photons are the main losses.

About 20% of the solar energy that reaches the earth's surface is not utilized by the most widespread silicon solar cells because photons with energies below the band-gap of silicon do not carry enough energy to generate free charge carriers. UC of these low energy photons is a promising approach to enhance the efficiency of solar cells. Detailed balance analyses indicate that the theoretical efficiency limit of silicon solar cells can be pushed from near 30% [11] to 40.2% via upconversion [12] for the illumination of one sun. For concentrated PV this limit can be enhanced even further. For a long time the host material sodium yttrium fluoride ($NaYF_4$) is know to provide highly efficient upconversion [13] For experimental realizations of silicon solar cells with upconversion, hexagonal sodium yttrium fluoride ($\beta$-$NaYF_4$) doped with trivalent erbium ($Er^{3+}$) [14], especially with a doping ratio of one erbium ion to four yttrium ions ($\beta$-$NaEr_{0.2}Y_{0.8}F_4$), has shown comparatively high quantum yield for UC of near infrared (NIR) photons at reasonably low excitation intensities for PV. An UC quantum yield of $\eta_{UC}$ = 4.3% was measured, at a monochromatic irradiance of 1370 W m$^{-2}$ and an excitation wavelength of 1523 nm, by photoluminescence measurements [15]. Due to the non-linearity of the UC processes the UC quantum yield increases with increasing irradiance [16,17]. An estimated luminescence UC quantum yield of 16.7% was determined by Richards et al. with an upconverter silicon solar cell device with laser illumination at 1523 nm and an irradiance of 24000 W m$^{-2}$ [18]. While these experiments can be considered as successful integration of upconversion into the PV, the achieved efficiencies at realistic irradiance levels are still too low, to make upconversion attractive for the application in commercial solar cells. Furthermore, the upconversion dynamics are not entirely understood yet, for example, why $\beta$-$NaYF_4$ is such a good host material and why small particles in the range of several 10 nm are less efficient than particles much larger as 0.1 µm [19].

The necessity of larger efficiencies and the lack of understanding create the need of detailed modeling of the upconversion dynamics, that allows to assess and to improve advanced concepts to increase upconversion efficiencies, such as the coupling with a second luminescent material [20-22] or the coupling with noble metal nanoparticles by utilizing plasmon resonance [23-26].

Up to now, extensive experimental and theoretical studies on the properties of different rare earth doped materials have mostly been carried out with regard to the use of these material systems in lasers or waveguide amplifiers. Therefore, the excitation wavelength was around 980 nm in contrast to approximately 1523 nm for UC applications in PV. The same applies for the development of simulation models [27-30]. In contrast, little modeling of the upconversion dynamics has been performed [31,32] in comparison to the large number of publications on the synthesis and characterization of new rare earth doped materials, such as Ref. [33-35]. No literature was found that addresses the special needs of the application in PV dealing with the material system of $\beta$-NaEr$_{0.2}$Y$_{0.8}$F$_4$ and no detailed investigations of modeling has been done concerning the optimization of the upconverter materials. However, the non-linear power dependence of UC were investigated by Pollnau et al [36] and Suyver et al. [37] by simplified rate equations and its solutions in the low and high power limit.

In this paper, we present a rate equation model to describe the nature of UC with all relevant processes. The rate equation model will be introduced in Sec. II. How the most important input parameters, the transition rates also known as Einstein coefficients, were determined from experimental data will be derived in Sec. III. In Sec. IV, we investigate the influence of various parameters for the ET processes and multi-phonon relaxation on the upconversion performance and discuss how reasonable the model describes the upconversion of $\beta$-NaEr$_{0.2}$Y$_{0.8}$F$_4$.

## II. THE RATE EQUATION MODEL

UC can occur by different processes [38]. The UC mechanisms with the highest relative efficiencies in rare earth based upconverters, like $\beta$-NaEr$_{0.2}$Y$_{0.8}$F$_4$, are ground state absorption (GSA), to populate an intermediate energy level with a comparable long lifetime in the range of ms, and subsequent energy transfer (ET) or excited state absorption (ESA) [39]. To learn more about the interaction of the different processes and to find possible leverages for optimization, we developed a simulation model based on rate equations. The model considers all significant processes, such as GSA and ESA, spontaneous emission (SPE), stimulated emission (STE), energy transfer (ET), and multi phonon relaxation (MPR). Only the lowest six energy levels of the Er$^{3+}$ are considered as the processes involving higher energy levels are of only minor relevance in the context of the application for upconversion solar cells. More than 95% of the observed UC luminescence stems from the transition from the energy level $^4I_{11/2}$ to the ground state with a peak wavelength of 980 nm, depending on the excitation intensity [15]. The $^4I_{11/2}$ energy level is populated by two photon processes. Hence the maximum achievable UC quantum yield would be 50%. Figure 1 visualizes the considered energy levels and processes of the UC model.

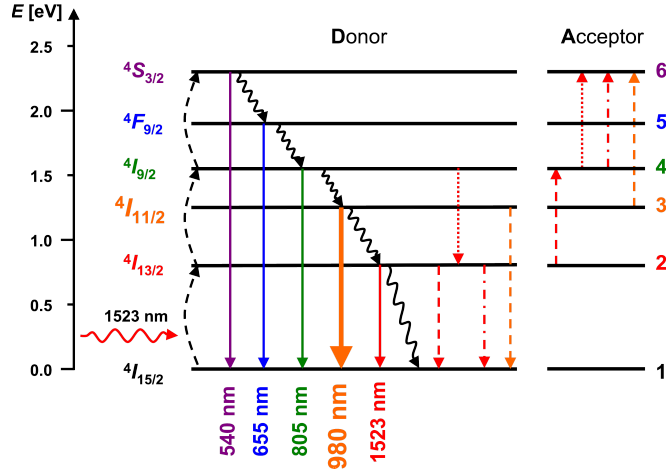

Figure 1:    Six-level upconversion model scheme of $Er^{3+}$. The energy levels correspond to the trivalent erbium in the host crystal $\beta$-NaYF$_4$ with corresponding luminescence wavelengths from the excited states to the ground state (solid arrows). The rate equation model assumes monochromatic excitation at 1523 nm. Higher energy levels can be populated by ground state absorption (GSA) followed by excited state absorption (ESA) (black dashed arrows). Another possibility to populate higher energy levels is energy transfer (ET) (pairs of dashed and dotted arrows), where energy is transferred from one ion, the donor (D), to a neighboring second ion, the acceptor (A). The four considered ET processes are labeled in the text with indices from 1 to 4, for the arrows from the left side to the right side. A distinction is made between upconversion energy transfer (ETU) processes, which is shown is this figure, and cross-relaxation (CR) processes. CR is the inverse process of ETU hence the arrows are turned around. Additionally, we consider multi-phonon relaxation (MPR) to next lower energy levels (waved arrows). The energy levels are denoted from 1 to 6 from the ground state to the highest considered energy level throughout the equations in this paper.

The occupation of the six considered energy levels is described by an occupation vector $\vec{n}$. The individual elements of the vector are the relative occupation of the specific energy level, i.e. the fraction of ions of a large ion ensemble that is excited to this state. In consequence, the norm of the occupation vector $\vec{n}$ is unity. The number of excited ions in a certain energy level $N_j$ can be derived by

$$N_j = n_j \rho_{Er} . \tag{1}$$

where $n_j$ is the relative occupation of a certain energy level $j$ and $\rho_{Er}$ is the density of erbium ions per unit volume in the host material, or in other words the volume density of $Er^{3+}$ ions. For simplification of the equations, throughout the paper the energy levels are numbered from 1 to 6 as for the energy levels from ground state $^4I_{15/2}$ to the highest considered energy level $^4S_{3/2}$. The occupation vector $\vec{n}$ and its rate of change $\dot{\vec{n}}$ are described by the differential equation:

$$\dot{\vec{n}} = \left[ M_{GSA} + M_{ESA} + M_{STE} + M_{SPE} + M_{MPR} \right] \vec{n} + \vec{v}_{ET}(\vec{n}).$$ (2)

The processes linear to the occupation vector $\vec{n}$ are described by matrixes. The matrix $M_{GSA}$ describes the ground state absorption, $M_{ESA}$ the excited state absorption, $M_{STE}$ the stimulated emission, $M_{SPE}$ the spontaneous emission and $M_{MPR}$ the multi-phonon relaxation. The energy transfer and cross relaxation processes $\vec{v}_{ET}(\vec{n})$ are not linear in $\vec{n}$ and therefore described by a set of vectors. The different processes and their mathematical representation are described in detail in the following subsections. The probabilities of all involved processes, except MPR, can be described by the Einstein coefficients for SPE. These coefficients have been determined from experimental data, which is discussed in Sec. III.

## A. Spontaneous emission

The probability of SPE is given directly by the Einstein coefficients $A_{if}$, which are based on the theory of photon-atom interaction [40]. The $A_{if}$ define the rate of radiative electron relaxation from an initial state $i$ to a final state $f$. The rate of the radiative transition is the inverse of the lifetime $\tau_i$ of the energy levels

$$A_i = \frac{1}{\tau_i} \quad \text{with} \quad A_i = \sum_f A_{if},$$ (3)

for the case where no other depopulation or population processes occur. The model considers SPE of transitions where strong luminescence is expected. Therefore, the matrix $SPE$ is written as:

$$M_{SPE} = \begin{pmatrix} 0 & A_{21} & A_{31} & A_{41} & A_{51} & A_{61} \\ 0 & -A_{21} & 0 & 0 & A_{52} & A_{62} \\ 0 & 0 & -A_{31} & 0 & 0 & 0 \\ 0 & 0 & 0 & -A_{41} & 0 & 0 \\ 0 & 0 & 0 & 0 & -A_{51}-A_{52} & 0 \\ 0 & 0 & 0 & 0 & 0 & -A_{61}-A_{62} \end{pmatrix}.$$ (4)

## B. Stimulated Processes

$M_{GSA}$, $M_{ESA}$ and $M_{STE}$ are the matrixes for the stimulated processes which can be described by the Einstein coefficient for SPE $A_{if}$ as well [40]. The probability for an absorption event $W_{abs,if}$ depends on the spectral photon energy density $u(\omega_{if})$

$$W_{abs,if} = \frac{\pi^2 c^3}{\hbar \omega_{if}^{\,3}} \frac{g_f}{g_i} u(\omega_{if}) A_{fi},$$

(5)

with $\omega_{if}$ being the angular frequency of the transition from level $i$ to $f$, $c$ the vacuum speed of light, $\hbar$ Planck's constant divided by $2\pi$, and $g_i$ and $g_f$ are the degeneracies of the energy levels $i$ and $f$ respectively. The degeneracies can be easily determined from the total angular momentum quantum number of the considered energy level $J_x$ via $g_x = (2J_x+1)$, where $x$ represents a certain energy level.

The simulation should be directly comparable to the photoluminescence experiments, in which the upconverter is excited by incident radiation with certain irradiance $I$. Therefore the formalism was adjusted to use the spectral irradiance $I_v(\omega)$ instead of the spectral energy density $u(\omega_{if})$ using the relation

$$I_v(\omega)\, d\omega = \frac{c}{n} u(\omega)\, d\omega,$$

(6)

with the frequency interval $d\omega$ and the refractive index $n$ of the upconverter. The model assumes monochromatic excitation with the angular frequency $\omega_{12}$, which corresponds to a wavelength of 1523 nm. The highest upconversion efficiency was experimentally found for this wavelength [15,41]. The probability of GSA and ESA can be described by the following combined matrix:

$$M_{GSA} + M_{ESA} = \frac{\pi^2 c^2 n}{\hbar \omega_{12}^{\,3}} I_v(\omega) \begin{pmatrix} -\gamma_1 \dfrac{g_2}{g_1} A_{21} & 0 & 0 & 0 & 0 & 0 \\ \gamma_1 \dfrac{g_2}{g_1} A_{21} & -\gamma_2 \dfrac{g_4}{g_2} A_{42} & 0 & 0 & 0 & 0 \\ 0 & 0 & 0 & 0 & 0 & 0 \\ 0 & \gamma_2 \dfrac{g_4}{g_2} A_{42} & 0 & -\gamma_3 \dfrac{g_6}{g_4} A_{64} & 0 & 0 \\ 0 & 0 & 0 & 0 & 0 & 0 \\ 0 & 0 & 0 & \gamma_3 \dfrac{g_6}{g_4} A_{64} & 0 & 0 \end{pmatrix}.$$

(7)

In reality, the transitions have a certain line width, which is not necessarily centered at 1523 nm. To take this into account, we introduced the damping factors $\gamma_1$, $\gamma_2$ and $\gamma_3$ representing the lower probability of the stimulated processes due to the energy mismatch between the transitions.

The inverse process of the stimulated absorption is STE. From equation (7) and the relation of the Einstein coefficients it can be simply derived that the probability for the STE is described via

$$M_{\mathrm{STE}} = \frac{\pi^2 c^2 n}{\hbar \omega_{12}^{\ 3}} I_\nu(\omega) \begin{pmatrix} 0 & \gamma_1 A_{21} & 0 & 0 & 0 & 0 \\ 0 & -\gamma_1 A_{21} & 0 & \gamma_2 A_{42} & 0 & 0 \\ 0 & 0 & 0 & 0 & 0 & 0 \\ 0 & 0 & 0 & -\gamma_2 A_{42} & 0 & \gamma_3 A_{64} \\ 0 & 0 & 0 & 0 & 0 & 0 \\ 0 & 0 & 0 & 0 & 0 & -\gamma_3 A_{64} \end{pmatrix}. \tag{8}$$

## C. Multi-phonon relaxation

Excited electrons of rare earth ions can also relax non-radiatively to lower energy levels by MPR. The transition probability depends on the phonon energies of the host crystal and the energy gap between the energy levels $\Delta E_{if}$. The probability for MPR $W_{\mathrm{MPR},if}$ was found empirically to follow approximately the relation [42,43]

$$W_{\mathrm{MPR},if} = W_{\mathrm{MPR}} \cdot e^{-\kappa \cdot \Delta E_{if}}. \tag{9}$$

$W_{\mathrm{MPR}}$ and $\kappa$ are material constants which depend on the host material that surrounds the optically active ions [3]. This relation is also called the energy-gap law. In general, low phonon energies of the host crystal reduce the losses due to MPR, because in that case many phonons are needed to bridge the energy gaps. Therefore, only neighboring energy levels are considered in the model. Accordingly, the matrix describing the probability for MPR is:

$$M_{\mathrm{MPR}} = W_{\mathrm{MPR}} \cdot \begin{pmatrix} 0 & +e^{(-\kappa\hbar\omega_{21})} & 0 & 0 & 0 & 0 \\ 0 & -e^{(-\kappa\hbar\omega_{21})} & +e^{(-\kappa\hbar\omega_{32})} & 0 & 0 & 0 \\ 0 & 0 & -e^{(-\kappa\hbar\omega_{32})} & +e^{(-\kappa\hbar\omega_{43})} & 0 & 0 \\ 0 & 0 & 0 & -e^{(-\kappa\hbar\omega_{43})} & +e^{(-\kappa\hbar\omega_{54})} & 0 \\ 0 & 0 & 0 & 0 & -e^{(-\kappa\hbar\omega_{54})} & +e^{(-\kappa\hbar\omega_{65})} \\ 0 & 0 & 0 & 0 & 0 & -e^{(-\kappa\hbar\omega_{65})} \end{pmatrix}. \tag{10}$$

## D. Energy transfer

ET is the dominant upconversion mechanism [39]. The energy is transferred from one excited ion, the donor, to another ion, the acceptor. Figure 2 visualizes this process and shows schematically the important parameters. A theoretical description of ET was developed by Förster [44] and subsequent works [45-47]. ET can be described as a function of the occupation of initial energy levels $n_{\mathrm{D},i}$ and $n_{\mathrm{A},j}$ from donor and acceptor ions. Most important parameters for the probability of the ET are the distance between these ions $d$, which corresponds to the erbium doping of the host crystal $\beta$-NaYF$_4$, and the overlap integral $k_{ET}$ of the normalized line shape functions $g_D(\omega)$ and $g_A(\omega)$, of the transitions within the donor and the acceptor:

$$k_{ET} = \int \frac{1}{\hbar} g_D(\omega) g_A(\omega) \, d\omega. \tag{11}$$

Figure 1 indicates which transitions match sufficiently to enable efficient ET and therefore are considered in the model. ET processes may also be assisted by phonons. This results in a possibly larger overlap integral of the involved transitions of acceptor and donor ion. However, the probability for a phonon assisted ET process decreases with the number of phonons required, similar to the energy gap law for multi-phonon relaxation [48,49].

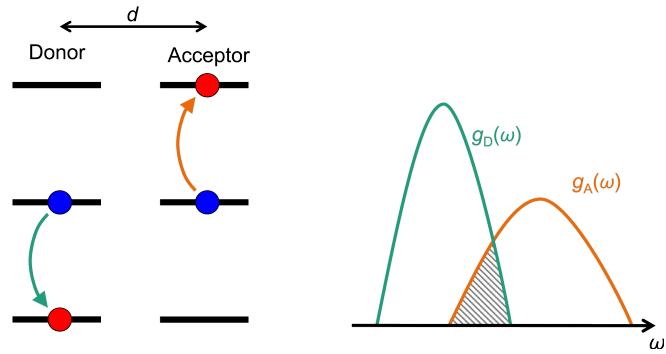

Figure 2: ET between a donor ion and an acceptor ion. The probability for ET depends on the distance $d$ of the two involved ions and the overlap (left) of the normalized line shape functions $g_D(\omega)$ and $g_A(\omega)$ (right).

The four considered ET processes of the model are labeled with indices from 1 to 4, for the arrows from the left side to the right side in Figure 1. A distinction is made between upconversion energy transfer (ETU) processes, which are shown in Figure 1, and cross-relaxation (CR) processes, which are the inverse processes to ETU. CR is an important mechanism for concentration quenching which reduces the quantum efficiency of UC and can therefore be considered as a loss mechanism.. However, CR might also populate intermediate states which might influence the UC positively. For example, if the electron of the donor ion relaxes from $^4S_{3/2}$ to $^4I_{11/2}$ the electron of the acceptor ion is excited from the ground state $^4I_{15/2}$ to $^4I_{11/2}$. Therefore, via CR suitable UC energy levels can be populated from one donor ion in the ground state and an acceptor ion in a highly excited state.

The probability for ET due to dipole-dipole interaction is [38]

$$W_{dd,ET} = \frac{27\hbar}{4} \frac{\pi c^6}{\omega_a^6} \frac{1}{n(n^2+2)^4} \frac{1}{d^6} A_{D,if} A_{A,jk} k_{ET} \tag{12}$$

with the average angular frequency $\omega_a = (\omega_{if} + \omega_{jk}) / 2$ of the transitions of the donor, from energy level $i$ to $f$, and acceptor, from energy level $j$ to $k$, and the Einstein coefficients for SPE of donor $A_{D,if}$ and acceptor $A_{A,jk}$. The rate by which a certain ET process occurs is consequently the product of the probability $W_{dd,ET}$ and the occupation of the starting energy levels $n_i$ and $n_j$. Therefore, the vector for ET is

$$\vec{v}_{ET}^{\,if,jk}(\vec{n}) = W_{dd,ET}\, n_i \vec{e}_{D,i}\, n_j \vec{e}_{A,j}\,.$$  (13)

with the corresponding unit vectors $\vec{e}_{D,i}$ and $\vec{e}_{A,j}$. By choosing the initial energy levels and the corresponding unit vectors one can decide whether it is the ETU or CR. In this work we use the following abbreviations:

$$ET_1 = \vec{v}_{ET}^{\,21,24}(\vec{n}) + \vec{v}_{ET}^{\,12,42}(\vec{n}) \qquad ET_2 = \vec{v}_{ET}^{\,42,46}(\vec{n}) + \vec{v}_{ET}^{\,24,64}(\vec{n})$$

$$ET_3 = \vec{v}_{ET}^{\,21,46}(\vec{n}) + \vec{v}_{ET}^{\,12,64}(\vec{n}) \qquad ET_4 = \vec{v}_{ET}^{\,31,36}(\vec{n}) + \vec{v}_{ET}^{\,13,63}(\vec{n})$$

The first term describes ETU, while the second term with permuted indices describes CR processes.

The vector $\vec{v}_{ET}(\vec{n})$ describes all ETU and CR processes considered in the model

$$\vec{v}_{ET}(\vec{n}) = ET_1 + ET_2 + ET_3 + ET_4\,.$$  (14)

In this model, we do not consider three ion interactions and assume the ions only interact with their direct neighbors. This is a reasonable assumption because the probability drops significantly with distance $d$ to the power of 6. In addition, we exclude the two crystal positions of the rare earth ions in $\beta$-NaYF$_4$ [50]. This means that only a certain distance between the ions is considered.

E.   General analysis assumptions

We are interested in the emission of photons that can generate free charge carriers in silicon. In our model, these emissions occur by transitions from the energy levels $^4I_{13/2}$ - $^4S_{3/2}$ to the ground state $^4I_{15/2}$ and from level $^4S_{3/2}$ to $^4I_{11/2}$. We neglect the possibility that emission from such levels is stimulated by a photon, as the intensities emitted and the occupation of these higher levels are relatively low which makes these processes very unlikely. The strength of upconversion luminescence $Lum_{if}$ from a higher energy level $i$ to the lower energy level $f$ is proportional to the luminescence rate $L_{if}$ which is calculated from the fraction of ions in a certain energy level $n_i$ and the spontaneous transition probability $A_{if}$

$$Lum_{if} \propto n_i A_{if} = L_{if}$$  (15)

The internal upconversion quantum yield $\eta_{Sim}$ of a certain transition is defined as the number of photons emitted from an excited energy level by SPE divided by the total number of absorbed photons, which are the sum of GSA and ESA rate subtracted by the STE rate:

$$\eta_{Sim} = \frac{n_i A_{if}}{\sum_{i \neq f} n_i M_{GSA} + n_i M_{ESA} - n_f M_{STE}} \,. \tag{16}$$

The overall quantum yield for upconverted photons emitted with energies above the band-gap of silicon can be calculated by summation over all involved $\eta_{Sim}$:

$$\eta_{Sim,UC} = \frac{\sum_{i=3}^{6} n_i A_{i1} + n_6 A_{62}}{\sum_{i \neq f} n_i M_{GSA} + n_i M_{ESA} - n_f M_{STE}} \,\hat{=}\, \eta_{UC} \,. \tag{17}$$

This quantity should be comparable with the experimentally determined upconversion quantum yield $\eta_{UC}$ from Ref. [15].

## III.   DETERMINATION OF MODEL PARAMETERS

The rate equation model presented in the previous section needs many input parameters. For an adequate UC simulation the parameterization is essential. In this section the determination of important input parameters will be described. The focus of this investigation will be on the Einstein coefficients. They are the fundamental input parameters of the rate equation model as they describe the transition probabilities of all individual processes, except MPR.

### A.   Einstein coefficients

The Einstein coefficients of the rare earth ions can be linked to the absorption coefficient of the material system [38]. We follow the approach to first determine the absorption coefficient related to the transitions from the ground state. Afterwards this data will be used to determine the Judd-Ofelt parameters [51,52], from which all relevant Einstein coefficients are calculated.

The material system $\beta$-NaEr$_{0.2}$Y$_{0.8}$F$_4$ is a microcrystalline powder, with a quite inhomogeneous size distribution. Due to the strong scattering in such materials the optical path length is not precisely known and it is not possible to determine the absorption coefficient directly from absorption measurements using the well known Beer-Lambert law. Instead we used the Kubelka-Munk theory [53,54] to derive the absorption coefficient $\alpha(\lambda)$ from measurements of the diffuse reflection spectrum from powder samples of $\beta$-NaEr$_{0.2}$Y$_{0.8}$F$_4$ with different thicknesses. The Kubelka-Munk (KM) theory is based on balance equations that describe the change in the irradiance while traversing an infinitesimally thin layer of material with a KM absorption coefficient $K(\lambda)$ and a KM scattering coefficient $S(\lambda)$. In order to determine $K(\lambda)$ two reflection measurements are necessary: one of an "infinitely" thick sample $R_\infty$, where no light is transmitted through the sample, and another one for a thin sample $R_t$, where some light is transmitted. The KM absorption coefficient is calculated by

$$K(\lambda) = \left( \frac{1}{2d_t} \right) \cdot \frac{1 - R_\infty(\lambda)}{1 + R_\infty(\lambda)} \cdot \ln \left( \frac{R_\infty(\lambda)\left(1 - R_t(\lambda)R_\infty(\lambda)\right)}{R_\infty(\lambda) - R_t(\lambda)} \right). \tag{18}$$

For the thick sample, the $\beta$-NaEr$_{0.2}$Y$_{0.8}$F$_4$ was filled into a powder cell. In this powder cell, the upconverter is compressed against a glass window of high transparency. The thickness of the powder layer is roughly 5 mm. To measure the reflection of thin layers, three samples were prepared. The powder was compressed between two glass slides. The thicknesses $d_t$ of the powder layers were determined to be 213 µm, 224 µm and 249 µm with an uncertainty for every sample of more than 5 %. The total reflection spectra of the samples were measured using a spectrophotometer Cary i500 with an integrating sphere and contain the direct and the diffuse reflection. The reflection of the glass slides were taken into account by reference measurements and subtracted from the reflection measurements of the UC samples. The absorption coefficient of the material $\alpha(\lambda)$ is proportional to $K(\lambda)$ and was calculated using the relation from Yang et al.[55] for diffuse light distribution

$$\alpha(\lambda) = K(\lambda)/(2\mu(\lambda)) = \begin{cases} K(\lambda)/(2S(\lambda)/K(\lambda))^{1/2}, & (2S(\lambda) \geq K(\lambda)) \\ K(\lambda)/2, & otherwise \end{cases}, \tag{19}$$

where $\mu(\lambda)$ is the scattering induced path variation factor, also called SIPV factor. Figure 3 shows the average absorption coefficient $\alpha(\lambda)$ determined from the three data sets created from the measurements on the three thin samples. The standard deviation of the three absorption spectra is around 11%. The average data was used to calculate the Einstein coefficients.

Since the absorption coefficient is a fundamental quantity for the determination of the Einstein coefficients the precision of the method was investigated. Therefore, the absorption coefficient of a glass filter and an optical glass was determined with the Beer-Lambert law. Afterwards, the glasses were grinded to a powder, the reflection of thick and thin powder samples were measured and the KM analysis was applied on the data. Comparisons of the two methods reveal a difference below ±15 %. However, strong absorption peaks contain larger errors of around ±20% because the light is absorbed on a much shorter path through the sample and the difference between the very thick and the thin sample is not sufficiently pronounced. This leads to a lower absorption coefficient with the KM method compared to the Beer-Lambert law. Therefore, the precision of the KM method will improved for thinner samples due to a larger difference of the reflection values. Further investigations on the validity and the precision of the KM analysis can be found in Ref. [55,56][57].

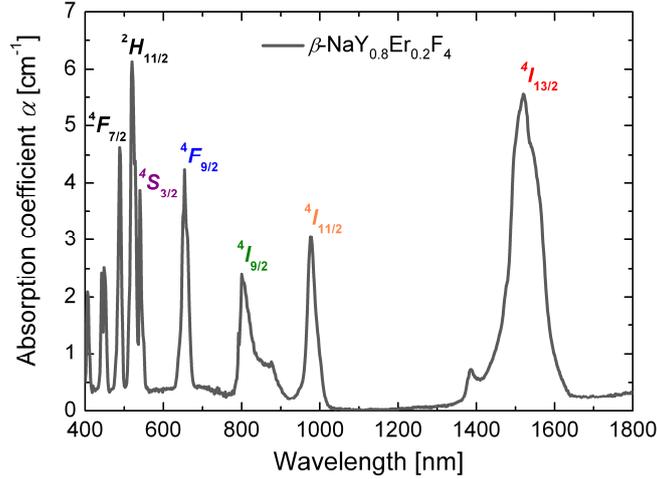

Figure 3:    Average absorption coefficient $\alpha(\lambda)$ of the $\beta$-NaEr$_{0.2}$Y$_{0.8}$F$_4$, calculated from the reflection spectrums of one thick and three thin samples with the Kubelka-Munk theory.

To calculate the Einstein coefficients, the transition probabilities must be linked to the absorption coefficient. The probability for SPE via an electric dipole process is given by

$$A_{if} = \frac{\omega_{if}^3 \chi}{3\pi \hbar c^3 \varepsilon_0 (2J_i + 1)} \left| \mu_{if} \right|^2 \tag{20}$$

where $\chi$ is the correction factor for the electric field in crystals with high symmetry [38], $\varepsilon_0$ the dielectric constant of the vacuum and $\left| \mu_{if} \right|^2$ the quadratic expected value of the dipole operator. This expected value of the dipole operator is linked to the absorption coefficient [38] via

$$\left| \mu_{if} \right|^2 = \frac{27 c \varepsilon_0 \hbar n (2J_i + 1)}{\omega_{if} \pi (n^2 + 2)^2 \rho_{CC}} \int \alpha(\omega) d\omega \tag{21}$$

With equation (20) and (21) it is possible to calculate the Einstein coefficients from the absorption data of the transitions from the ground state to excited states. For the UC model, however, also the coefficients for transitions between excited energy levels are necessary. Due to the theory of Judd [51] and Ofelt [52]. the probability of all transitions can be calculated from the absorption coefficient.

Table I:  Einstein coefficients $A_{if}$ for SPE of the $\beta$-NaEr$_{0.2}$Y$_{0.8}$F$_4$ determined by Judd-Ofelt analysis. The columns

indicate the level from which the emission occurs ($i$) and the rows the final state after the emission ($f$).

| [s$^{-1}$] | $^4I_{13/2}$ | $^4I_{11/2}$ | $^4I_{9/2}$ | $^4F_{9/2}$ | $^4S_{3/2}$ |
|---|---|---|---|---|---|
| $^4I_{15/2}$ | 122.1 | 153.1 | 22.5 | 766.9 | 1451.6 |
| $^4I_{13/2}$ | | 7.0 | 2.4 | 142.4 | 390.0 |
| $^4I_{11/2}$ | | | 0.1 | 28.0 | 130.5 |
| $^4I_{9/2}$ | | | | 5.0 | 51.2 |
| $^4F_{9/2}$ | | | | | 7.7 |

Following this theory, the expected value of the dipole operator can be calculated for transitions within

the 4f shell of rare earth ions via

$$\left| \mu_{if} \right|^2 = e^2 \sum_{t=2,4,6} \Omega_t \cdot \left| \left\langle f \, J_f \left\| U^{(t)} \right\| i \, J_i \right\rangle \right|^2 . \tag{22}$$

In this equation, the two involved states are identified by the quantum numbers $i$ and $f$ with corresponding total angular momentum $J_f$ and $J_i$, respectively. $U^{(t)}$ is a tensor operator of rank t.

Reduced matrix elements of this tensor operator can be found in literature for different host crystals. In this work, we used the values from Carnall for Er$^{3+}$ in a LaF$_3$ host crystal [58] which is a reasonable approximation for the situation of a $\beta$-NaYF$_4$ host crystal, because the materials are very similar. From the reduced matrix element of a certain transition a first impression which transitions are likely to be strong or weak can be estimated. The $\Omega_t$ are the Judd-Ofelt intensity parameters. They characterize the strength of the crystal field. With least square fits of equation (21) and (22) to the experimentally determined  absorption  data,  the  values  of  $\Omega_t$  were  determined  to  be  $\Omega_2 = 1.4 \times 10^{-24}$ m$^2$,

$\Omega_4 = 2.0 \times 10^{-25}$ m$^2$ and $\Omega_6 = 1.9 \times 10^{-24}$ m$^2$. With these Judd-Ofelt parameters and the literature values of

the reduced matrix elements all transition probabilities were calculated:

$$A_{if} = \frac{\omega_{if}^3 \chi}{3\pi \hbar c^3 \varepsilon_0 (2J_i+1)} \, e^2 \sum_{t=2,4,6} \Omega_t \left| \left\langle f \, J_f \left\| U^{(t)} \right\| i \, J_i \right\rangle \right|^2 \tag{23}$$

The results are summarized in Table I. In general, the obtained values reflect the fact that SPE is more pronounced for transitions bridging a larger energy gap. For the transitions from $^4I_{9/2}$, however, small values were obtained. This is also reflected in the reduced matrix elements for this transition which are very low and actually zero for $U^{(2)}$. However, considering that a quite complex combination of theories, all involving certain approximations and estimations, was necessary to calculate these parameters, these results must be considered rather as an approximation than as exact values. Nevertheless, the values are adequate to describe the UC reasonable (see Sec IV).

## B.   Other parameters

The model requires several other input parameters: Some could be taken directly from literature others had to be determined experimentally. Table II lists all final input parameters after the optimization described in Sec. IV.

We used a refractive index of $n = 1.5$ in our calculations, based on the work of Thoma [59] from similar fluoride materials. To calculate the average distance between neighboring ions the density of the $\beta$-NaEr$_{0.2}$Y$_{0.8}$F$_4$ powder was determined experimentally by the water displacement method. Based on the assumption of a powder volume fill factor of 0.5 in air when dry [60], a density of the powder of $(4.7 \pm 0.5)$ g cm$^{-3}$ was calculated. With the given crystal structure [50] and under the assumption of a homogeneous distribution of the ions in the medium this results in a density of the absorbing Er$^{3+}$ of $\rho_{Er} = 1.4 \times 10^{21}$ cm$^{-3}$, which corresponds to an average distance between the Er$^{3+}$ of $d = 0.9$ nm. This value is in fairly good agreement with values from literature for single crystals of comparable material systems, of which the lattice structure is well known [14,50,61].

Other parameters where optimized to achieve good agreement between simulation and experiment. Starting values for the optimization were estimated based on literature: The parameters for MPR, $W_{MPR}$ and $\kappa$, were determined by fits of experimental data of comparable host materials from Weber [3,62] with equation (9) considering that the average of the dominant phonon modes is 360 cm$^{-1}$ for $\beta$-NaYF$_4$ [63]. With this method the initial values for $W_{MPR}$ and $\kappa$ were determined to be $W_{MPR} = 10^8$ s$^{-1}$ and $\kappa = 2.32 \times 10^{20}$ J$^{-1}$. These parameters were further optimized in the simulations, as described in section IV.

The four overlap integrals $k_{ET,1}$ to $k_{ET,4}$ and the damping factors $\gamma_1$, $\gamma_2$ and $\gamma_3$ were treated as free parameters of the model. The order of magnitude of the $k_{ET}$ was taken from Henderson [38] and the starting values were first estimated by taking the different energy mismatch of the involved transitions into account. The damping factors were estimated from the difference of the transition wavelength to the excitation wavelength and the expected line width of the transition. For these estimations, energy mismatch and spectral width were determined based on the absorption data presented in Figure 3.

Table II: Standard (final) input parameters for the rate equation model. These values are used for the red solid lines in the graphs of Sec. IV.

| Parameter | Value | Unit | Description | Source |
|---|---|---|---|---|
| $\kappa$ | $2.15 \times 10^{20}$ | $J^{-1}$ | Constant for MPR, depends on the host crystal | Literature[#] |
| $n$ | 1.5 | | Refractive index of $\beta$-NaY$_{0.8}$Er$_{0.2}$F$_4$ | Literature |
| $W_{MPR}$ | $1.0 \times 10^8$ | $S^{-1}$ | Constant for MPR depends on the host crystal | Literature[#] |
| $\Delta\omega$ | $8.1 \times 10^{11}$ | $S^{-1}$ | Spectral frequency interval of the illumination | Experimental |
| $\rho_{ER}$ | $1.4 \times 10^{21}$ | $cm^{-3}$ | Density of erbium ions | Experimental |
| $d$ | 0.9 | nm | Average distance between the Er$^{3+}$ in the host crystal | Experimental |
| $I_\nu(\omega), I$ | variable | $W\,m^{-2}$ | Spectral irradiance on the UC, also called irradiance when divided by the spectral frequency interval of the illumination | Experimental |
| $\gamma_1$ | 1.0 | | Damping factor for the energy mismatch between excitation and transition energy for the transition of $^4I_{15/2}$ to $^4I_{13/2}$ | Free parameter* |
| $\gamma_2$ | 0.5 | | Damping factor for the energy mismatch between excitation and transition energy for the transition of $^4I_{15/2}$ to $^4I_{9/2}$ | Free parameter* |
| $\gamma_3$ | 0.8 | | Damping factor for the energy mismatch between excitation and transition energy for the transition of $^4I_{15/2}$ to $^4S_{3/2}$ | Free parameter* |
| $k_{ET,1}$ | $8.0 \times 10^{18}$ | $J^{-1}$ | | |
| $k_{ET,2}$ | $1.0 \times 10^{18}$ | $J^{-1}$ | Energy overlap integral of the transition from donor to acceptor | Free parameter* |
| $k_{ET,3}$ | $0.5 \times 10^{18}$ | $J^{-1}$ | | |
| $k_{ET,4}$ | $7.0 \times 10^{18}$ | $J^{-1}$ | | |

*Estimation with experimental or theoretical considerations.
[#]Calculated from literature values of similar materials.

## IV.  RATE EQUATION SIMULATIONS

The dynamics of the system are shown in Figure 4 for the considered energy levels of the upconverter at a continuous irradiance of 1000 W m$^{-2}$. The $L_i$ were calculated with equation (15). At the beginning, the ground state is completely filled and the other energy levels are empty. The simulation begins at $t$ = 0 s, and over time higher energy levels are occupied. The luminescence $L_i$ follows the occupation of the energy levels, because of the assumed linear dependence. After 25 ms the steady state, the balance between population and depopulation, is reached. ET processes depend on the occupation of several energy levels. Therefore, they are not significant from the beginning. For example, at $t$ = 0.3 ms a small bend in the $L_i$ of the $^4I_{11/2}$ occurs. This is most likely caused by the $ET_1$ process due to a larger occupation of $^4I_{13/2}$. For larger irradiance the stimulated processes, such as GSA or ESA, are much stronger and the dynamic of the system is much faster.

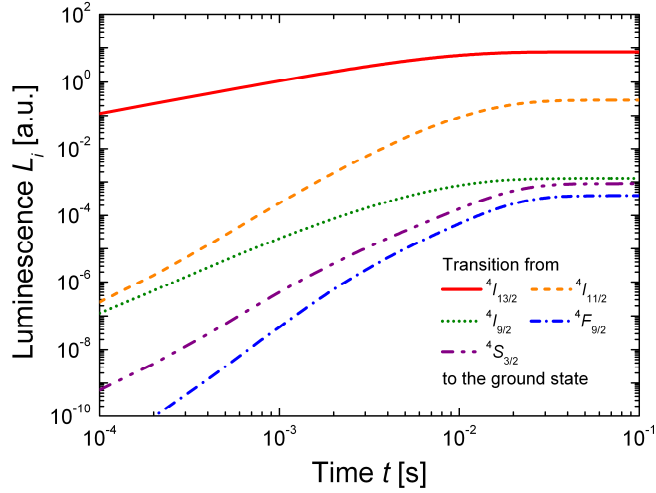

Figure 4: Dynamics of the luminescence Li for the considered energy levels of the upconverter for an irradiance of 1000 W m-2 on a double logarithmic scale. The Li follows the occupation of the energy levels linearly. At t = 0 s, all ions are in the ground state. The energy levels of the ions are occupied over time and luminescence occurs, due to spontaneous emission to the ground state. After 25 ms, the steady state is reached.

For the following evaluation of our model and the optimization of the parameters, we compared the experimentally determined upconversion quantum yield $\eta_{UC}$ of the $\beta$-NaEr$_{0.2}$Y$_{0.8}$F$_4$ with the $\eta_{Sim,UC}$ from equation (17) and the $\eta_{Sim}$ from equation (16). The simulated values were determined after 200 ms. Since, the effects of the parameters are interdependent of each other, the analysis of the optimization of the model parameters is complex and a methodical approach is necessary. In this section we will discuss the most important parameters for energy transfer, multi-phonon relaxation and the irradiance dependence of the system. First, the parameters have been varied at an irradiance of 1000 W m$^{-2}$ and the simulated quantum yield for the single transitions $\eta_{Sim}$ were optimized to fit the absolute $\eta_{UC}$ and the individual contributions of $\eta_{Sim}$ on the total quantum yield $\eta_{Sim,UC}$. Second, the irradiance dependence was calculated for selected values of the parameters. In these analyses, the irradiance $I$, should be understood to mean $I_v(\omega)$ divided by the spectral frequency interval of the illumination $\Delta\omega$. The standard values for the parameters of the UC simulations, also called final parameters in this work, are listed in Table II. These values are those that best fit the measurements. Details on the experimental analysis of $\beta$-NaEr$_{0.2}$Y$_{0.8}$F$_4$ can be found in Ref. 15.

## A.    Investigation on energy transfer

The influence of the four different ET processes on the internal quantum yield $\eta_{\text{Sim}}$ for the representative energy levels $^4I_{11/2}$ and $^4F_{9/2}$ are presented in Figure 5 and Figure 6 for an irradiance of $I = 1000$ W m$^{-2}$. Each of the $k_{ET}$ was varied individually over several orders of magnitude by keeping the others at the standard values.

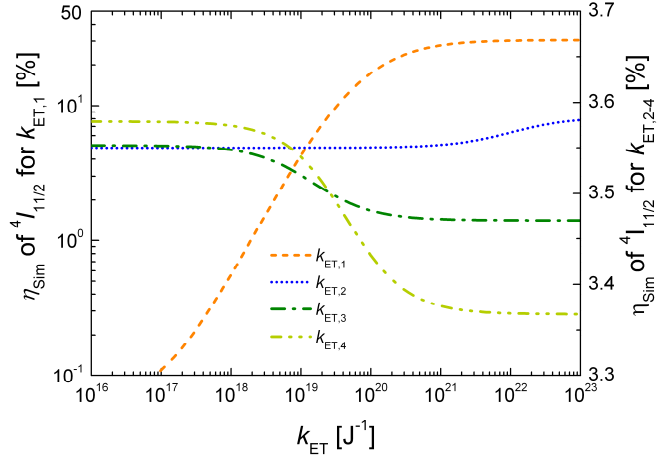

Figure 5:    Individual variations of the four overlap integrals $k_{\text{ET},1}$ to $k_{\text{ET},4}$ and their influence on the $\eta_{\text{Sim}}$ of the transition from $^4I_{11/2}$ to the ground state $^4I_{15/2}$ on a double logarithmic scale. The impact of $k_{\text{ET},1}$ is by far the strongest. The $\eta_{\text{Sim}}$ rises for increasing $k_{\text{ET},1}$ over several orders of magnitude. The $\eta_{\text{Sim}}$ for $k_{\text{ET},2}$ to $k_{\text{ET},4}$ are shown on the right scale, because the changes are much smaller. With large $k_{\text{ET},2}$, CR processes are stronger than UC ET and the $^4I_{9/2}$ is becoming more populated, and therefore the $\eta_{\text{Sim}}$ of $^4I_{11/2}$ increases indirectly due to MPR. On the other hand, with increasing $k_{\text{ET},3}$ and $k_{\text{ET},4}$ higher energy levels are more populated, which decreases the $\eta_{\text{Sim}}$ of $^4I_{11/2}$.

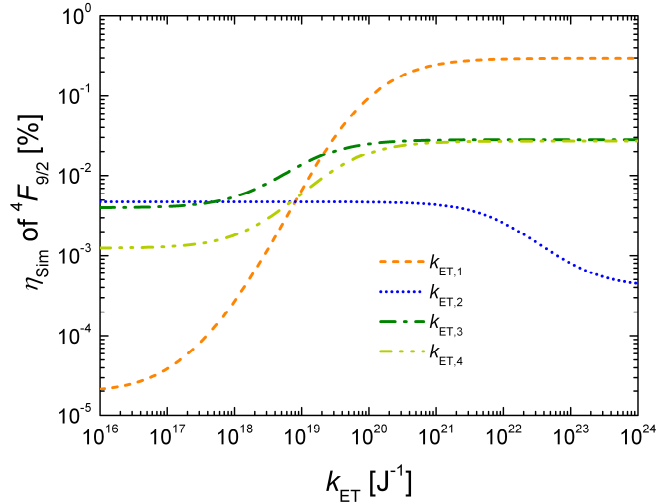

Figure 6:    Individual variations of the four overlap integrals of the four ET processes corresponding to Figure 5, but for the transition from $^4F_{9/2}$ to $^4I_{15/2}$. With increasing $k_{\text{ET},1}$, the energy levels $^4I_{9/2}$ and $^4I_{11/2}$ are more populated, which is also beneficial for the other ET processes, because the probability is proportional to the occupation of these energy levels. This also means that neighboring ions are more likely to be in suitable energy states for an ET process. The reasons for the dependence of various $k_{ET}$ on the $\eta_{Sim}$ of $^4F_{9/2}$ correspond to the $\eta_{Sim}$ of $^4I_{11/2}$.

The ETU process of $ET_1$ is assumed to be the most efficient process for UC. First, the overlap integral $k_{ET,1}$ is likely to be large, because of the broad transitions involved and the relatively high resonance of the involved transitions. Second, the occupation of the energy levels, especially $^4I_{15/2}$ and $^4I_{13/2}$, are particularly high, hence it is more likely that neighboring ions are in suitable energy states. The strong decreasing $\eta_{Sim}$ of $^4I_{11/2}$ for small $k_{ET,1}$ and the fact that by far most upconverted photons originate from this transition, shows the significance of this ETU process for the total UC quantum yield, shown in Figure 5. The influence of $k_{ET,1}$ on the $^4F_{9/2}$ level, shown in Figure 6, is mostly attributed to the larger occupation of the lower energy levels like $^4I_{11/2}$ and $^4I_{9/2}$ and corresponding larger probabilities for $ET_3$ and $ET_4$.

Since the occupation probabilities for the energy levels above $^4I_{13/2}$ is very low, the ETU processes of $ET_2$ to $ET_4$ only marginally affect the $\eta_{Sim}$ of $^4I_{11/2}$. The values of $\eta_{Sim}$ for these ET processes correspond to the scale on the right in Figure 5. For very large $k_{ET,2,}$ one can see that the CR process of $ET_2$ is stronger than ETU and the $^4I_{9/2}$ is becoming more populated. Hence the $\eta_{Sim}$ of $^4I_{11/2}$ increases indirectly due to MPR from $^4I_{9/2}$. For increasing $k_{ET,3}$ and $k_{ET,4,}$ higher energy levels are more populated, which decreases the $\eta_{Sim}$ of $^4I_{11/2}$. The higher energy levels are much more affected by these ET processes. Figure 6 shows how the $\eta_{Sim}$ of $^4F_{9/2}$ increases for larger values of $k_{ET,3}$ and $k_{ET,4}$, which is quite obvious, because the occupation of $^4F_{9/2}$ benefits from ETU processes from the respectively highly occupied energy levels $^4I_{13/2}$ and $^4I_{11/2}$. For very high values of $k_{ET,2}$, the $\eta_{Sim}$ of $^4F_{9/2}$ decreases, because CR is favored at this point. The low occupation of the higher energy levels in respect to the lower energy level is again the main reason for larger CR probability of $ET_2$, whereas the transitions are from $^4S_{3/2}$ to $^4I_{9/2}$ for the donor and from $^4I_{13/2}$ to $^4I_{9/2}$ for the acceptor.

The analysis above shows how a single ET process changes the individual $\eta_{Sim}$ of the upconverter. However, the dynamic of the model is complex and small changes of other parameters than the ones currently under consideration will change the overall dynamics and thus the results of the simulations. Therefore, we would like to point out the importance of correct parameterization, with physically reasonable input parameters, in obtaining sufficiently realistic simulations of upconversion.

How results will change if we change all ET processes simultaneously is very interesting, as changing the distance between the ions $d$ will alter all ET processes, which influences the $\eta_{Sim}$ of the various transitions, as shown in Figure 7. In comparison to Figure 5 and Figure 6, there is a peak at $d$ = 0.45 nm, which drops for shorter $d$. This means that there is a best suited doping ratio of the host material with $Er^{3+}$ to achieve a high quantum yield for an irradiance of 1000 W m$^{-2}$. For a different irradiance the situation may change (compare with Figure 8). A. The values for d < 0.6 nm, however, seem to be overestimated because very high $Er^{3+}$ content are necessary to achieve such distances between neighboring $Er^{3+}$ ions. Therefore the presented values for the distance should be taken with care.

In general, ET processes become proportionally stronger for an increasing occupation of higher energy levels for smaller $d$,. Therefore, the probability of CR processes will also increase and force the $\eta_{Sim}$ downwards. Especially CR of $ET_2$, which occupies $^4I_{9/2}$, contributes to this behavior, which is why $\eta_{Sim}$ of $^4I_{9/2}$ does not show a peak and increases continuously for smaller $d$.

For larger $d$, the $\eta_{Sim}$ of all energy levels drops drastically. The $\eta_{Sim}$ of the lower energy levels $^4I_{9/2}$ and $^4I_{11/2}$ have constant values for $d$ > 3 nm. At these distances, the probability for ET is negligible and the energy levels are mainly occupied by GSA and ESA, which do not depend on $d$.

The internal upconversion quantum yield $\eta_{Sim,UC}$ in dependence of the irradiance is shown in Figure 8 for different $d$. A good agreement between simulations and the experimentally determined upconversion quantum yield $\eta_{UC}$ (black squares for all figures) was found for $d$ = 0.9 nm. For smaller $d$, the ET processes are much stronger, which results in a much steeper rise and larger values of $\eta_{Sim,UC}$ at lower irradiance. However, the shape of the curves is more concave for smaller $d$ and saturation effects occur at much lower irradiance. For distances of 0.1 nm and 0.3 nm, saturation occurs at around 1200 W m$^{-2}$ and 600 W m$^{-2}$, respectively. Therefore, no enhancement of the upconversion quantum yield can be achieved by increasing the irradiance even further. As before, these results need to be taken with care.

Nevertheless, for applications in PV saturation effects at high irradiances can be advantageous because the irradiance expected from the sun in the suitable range for UC is roughly only 120 W m$^{-2}$. For low concentrated PV of around 10 suns an irradiance of 1000 W m$^{-2}$ is a reasonable estimation, when losses due to the concentration optics are considered. Therefore, high doping ratios may be beneficial for PV, because of the larger quantum yield at low irradiance. The irradiance dependent quantum yield curves for different $k$ intersect each other at certain irradiance, which means that for different light concentrations the best suited doping ratio changes.

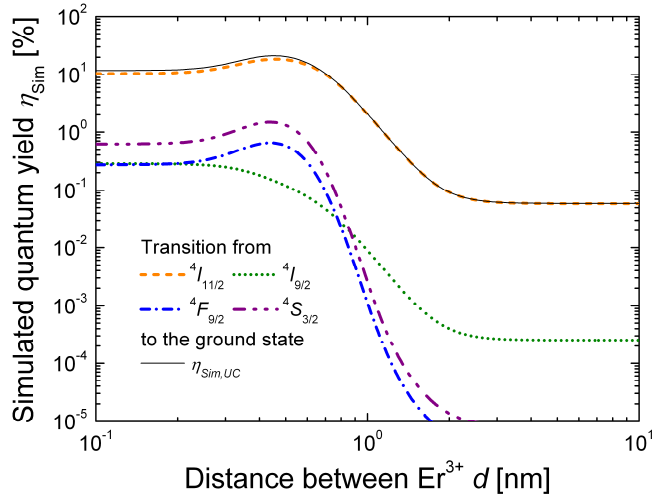

Figure 7:   The $\eta_{Sim}$ of the various transitions depends on the distance between neighboring ions $d$. Variation of $d$ corresponds to simultaneous modification of all ET processes. For large distances, the $\eta_{Sim}$ of $^4F_{9/2}$ and $^4S_{3/2}$ drops drastically. The $\eta_{Sim}$ for the lower energy levels $^4I_{9/2}$ and $^4I_{11/2}$ are at constant values for $d > 3$ nm. At these distances, virtually no ET occurs and the energy levels are occupied by GSA and ESA.

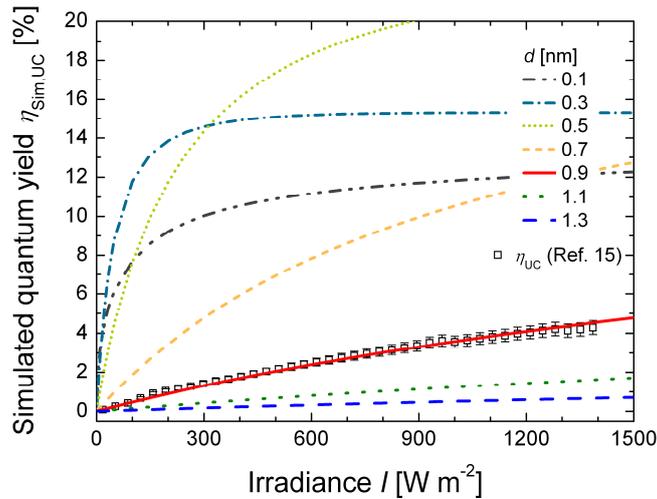

Figure 8:   Irradiance dependence of $\eta_{Sim,UC}$ for different $d$. For shorter $d$, ET processes are much stronger and larger values of $\eta_{Sim,UC}$ at lower irradiance can be reached. However, the shape of the curves is more concave for shorter $d$ and saturation effects occur at much lower irradiance. At a distance of 0.9 nm, the simulations agree best with the experimental data $\eta_{UC}$ (black squares). Results for $d < 0.6$ nm seem to be overestimated and should be taken with care.

B.    Investigations on multi-phonon relaxation

The lower the value of $W_{MPR}$, the longer is the lifetime of the energy levels. Therefore, the probability for upconversion is increasing for low non-radiative relaxation rates due to the longer lifetime of the intermediate energy levels. This behavior is reflected in the increasing values of $\eta_{Sim}$ for decreasing $W_{MPR}$, see Figure 9. The simulations show a peak at $4 \times 10^7$ s$^{-1}$ and $\eta_{Sim,UC}$ decreases slightly for lower $W_{MPR}$, because the $^4I_{11/2}$ mainly populated by MPR. For comparison, values of $W_{MPR}$ for various host materials of rare-earth can be found in Ref. [3,17]. A value of $3.7 \times 10^8$ s$^{-1}$ was determined for LaF$_3$, for example.

For small values of $W_{MPR}$, the $\eta_{Sim}$ of $^4I_{11/2}$ drops, because $^4I_{11/2}$ is less populated by MPR and the fraction of the depopulation by the ETU process of $ET_{4'}$ is increasing. A lower occupation of $^4I_{11/2}$, with a larger SPE rate compared to $^4I_{9/2}$, is the main reason for the decreasing $\eta_{Sim,UC}$ at small $W_{MPR}$.  The contribution of the transitions from $^4I_{9/2}$ and $^4S_{3/2}$ on the $\eta_{Sim,UC}$ increases for smaller $W_{MPR}$, due to smaller rates of MPR to lower energy levels.

However, three photons are needed to populate $^4S_{3/2}$, which reduces $\eta_{Sim,UC}$ as well. The dependence of $\eta_{Sim,UC}$ on the irradiance for different values of $W_{MPR}$ is shown in Figure 10. The $W_{MPR}$ does not modify the shape of the curves significantly. For smaller values of $W_{MPR}$, the curves are slightly less concave.

In general, there is a competition between the various processes. With increasing $W_{MPR}$ the relaxation of the systems is more and more done by the non-radiative processes. Consequently the fraction of the radiative processes on the total transition rates decrease as well as $\eta_{Sim,UC}$. In addition, the lifetime of the intermediate level shortens, which makes the occupation of higher energy level less likely. There is a different shape of the curve for $^4F_{9/2}$ with a peak at $W_{MPR} = 2 \times 10^8$ s$^{-1}$, see Figure 9. This energy level is only populated by MPR, and the energy gap to the higher energy level $^4S_{3/2}$ is lower than to the next lower energy level $^4I_{9/2}$. Therefore, the non-radiative transition rate for the population of $^4F_{9/2}$ is larger than the depopulation rate. In our simulation the occupation of $^4F_{9/2}$ would only be increasing, but due to the lower occupation of all higher energy levels the occupation decreases for large values of $W_{MPR}$. However, temperature-dependent lifetime measurements of $\beta$-NaEr$_{0.2}$Y$_{0.8}$F$_4$ indicated that the $^4F_{9/2}$ is not only occupied by MPR, because the lifetime did not change as significantly as expected [63]. Additionally, while the lifetime of $^4S_{3/2}$ decreases with increasing Er$^{3+}$ content the lifetime of $^4F_{9/2}$ changes only marginally. This was attributed to stronger CR of the $^4S_{3/2}$ for larger Er$^{3+}$ content in the host material and consequently shorter distances between the Er$^{3+}$. Further investigations have to consider this observation and an even more complex model will be necessary.

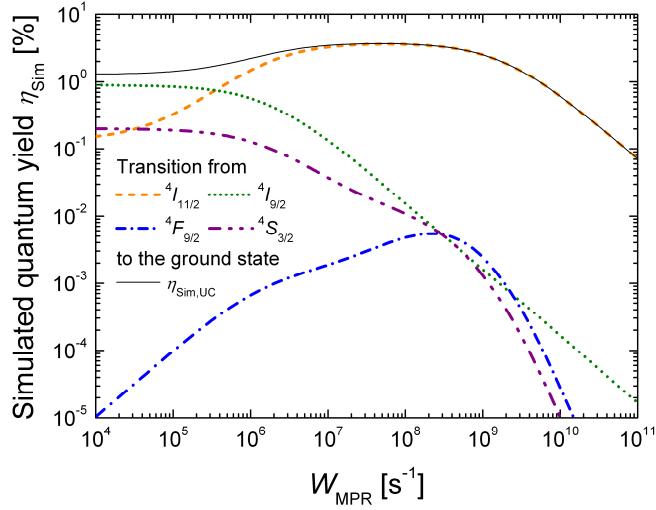

Figure 9: The $\eta_{Sim}$ of the energy levels drops for very high values of $W_{MPR}$, because the probability of MPR increases and becomes the dominant process. However, for lower $W_{MPR}$ it is obvious that the energy levels $^4I_{11/2}$ and $^4F_{9/2}$ are less populated, because the rate of MPR is decreasing and these levels are mainly occupied by this process. The simulation shows the best tradeoff between non-radiative losses and beneficial population by MPR for $W_{MPR} = 4 \times 10^7$ s$^{-1}$.

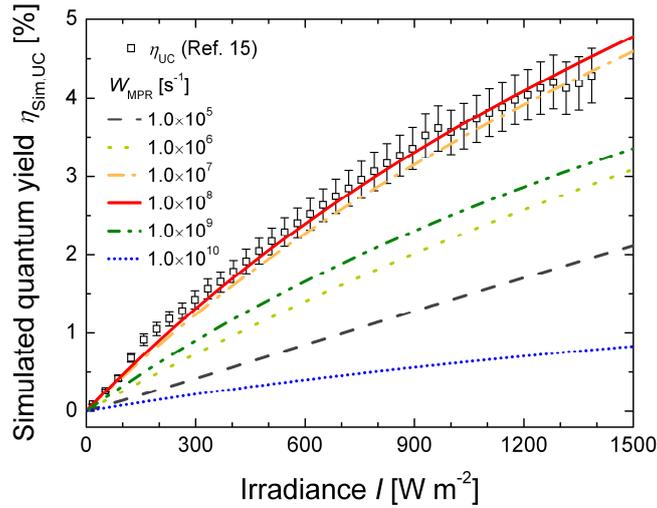

Figure 10: Irradiance dependence of $\eta_{Sim,UC}$ for different values of $W_{MPR}$. The black squares show the experimental curve of $\eta_{UC}$. For $W_{MPR} = 10^8$ s$^{-1}$ the experiment can be described nicely. The absolute values of $\eta_{Sim,UC}$ first increases with increasing $W_{MPR}$, mainly because the energy level $^4I_{11/2}$ is more populated and has a high SPE rate. After a maximum the $\eta_{Sim,UC}$ decreases with increasing $W_{MPR}$, because the MPR is becoming the dominant process. The shape of the irradiance dependence $\eta_{Sim,UC}$ is not affected significantly by changing $W_{MPR}$.

The $\kappa$ from equation 9 is related to the number of phonons needed to bridge the energy gap between the energy levels. Therefore, with $\kappa$ the individual MPR rates can be tuned due to the exponential dependence. In consequence, $\kappa$ changes the fraction of $\eta_{Sim}$ for the different transitions on the $\eta_{Sim,UC}$, while $W_{MPR}$ modifies all non-radiative transitions simultaneously.

Figure 11 shows the dependence of $\kappa$ on the transitions of the different energy levels. The $\eta_{Sim}$ are very sensitive to small changes of $\kappa$. With increasing $\kappa$ the difference between the relaxation rates decreases, because the exponential curve is quite flat. In addition, the probability of MPR decreases and reduces the occupation of $^4I_{11/2}$ and $^4F_{9/2}$. Therefore, these energy levels contribute less to the $\eta_{Sim,UC}$, similarly to the discussion on $W_{MPR}$ before.

For decreasing $\kappa$, the difference between the MPR rates of the different transitions increases. However, for $\kappa < 2.3 \times 10^{20}$ J$^{-1}$ the $\eta_{Sim,UC}$ decreases, because MPR becomes the dominant process.

The simulations show a maximum of $\eta_{Sim,UC}$ for $\kappa = 2.3 \times 10^{20}$ J$^{-1}$. As with $W_{MPR,}$ the most suitable balance between non-radiative losses and population of $^4I_{11/2}$ is the main reason for the high UC quantum yield. The shape of the irradiance dependent $\eta_{Sim,UC}$ for different input parameters of $\kappa$ does not change significantly, see Figure 12.

For all processes, one has to find the optimal parameters to make the UC material most efficient. For example, lower MPR probabilities do not automatically enhance the UC quantum yield. In order to achieve efficient UC the meta-stable energy level $^4I_{11/2}$ need to be highly populated, because SPE is the dominant depopulation process compared to MPR and ETU process of $ET_4$

For a long time NaYF$_4$ is known for one of the most efficient hosts for rare-earth based upconverters. Up to now, the low phonon energy is one of the main arguments for this issue. If the phonon energy of the host material is too low the UC will not be highly efficient and vise versa. The first reason is the above mentioned occupation of intermediate energy levels with large SPE rate and small depopulation rate. Second, from our data no ET process was found to be in complete resonance and consequently the overlap integral from equation 11 would be negligibly small. Therefore, the majority of ET processes seem to be phonon assisted and the probability for these ET process is also depending on the number of phonons needed to bridge the energy gap, as described in Sec. II.D. Hence, there will be ideal phonon energy between the non-radiative losses and phonon assisted ET. This possibly explains why usually upconverter based on bromide or chloride hosts with lower phonon energies than fluoride hosts are less efficient host materials for NIR to VIS upconversion.

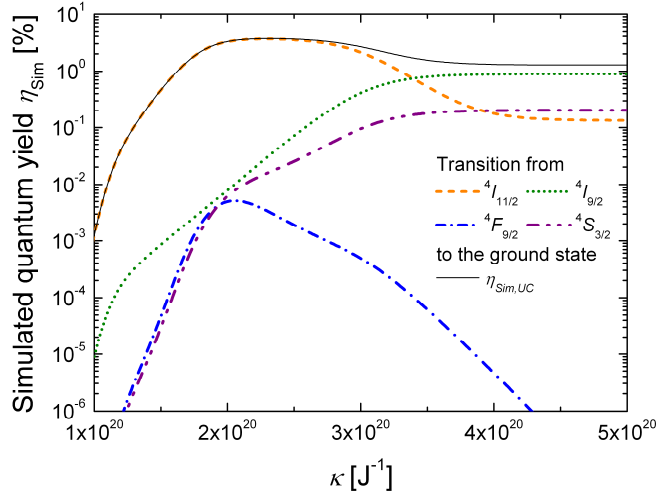

Figure 11: The $\eta_{Sim}$ of the different energy levels are very sensitive to small changes of $\kappa$. With $\kappa$ the MPR rate between transitions with different energy gaps can be tuned, which modifies the fraction of $\eta_{Sim}$ for the different transitions. For larger $\kappa$, the probability of MPR decreases and reduces the occupation of $^4I_{11/2}$ and $^4F_{9/2}$. Therefore, these energy levels contribute less to the $\eta_{Sim,UC}$. For $\kappa < 2.3 \times 10^{20}$ J$^{-1}$ all $\eta_{Sim}$ decrease, because the MPR rate is rising and becomes the dominant process.

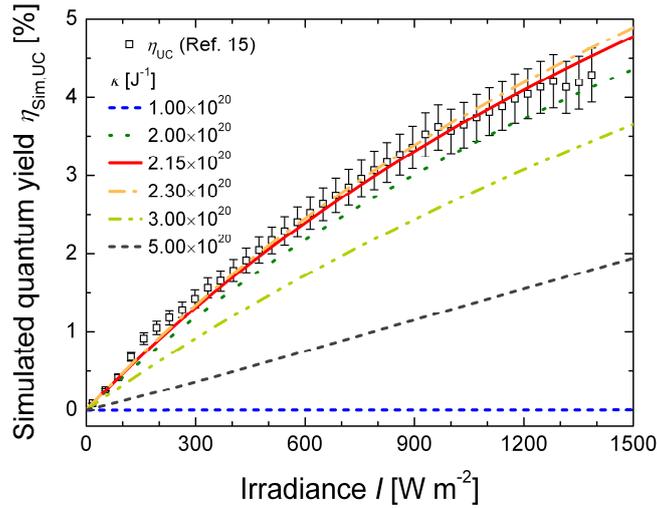

Figure 12: The $\eta_{Sim,UC}$ for different input parameters of $\kappa$ in dependence of the irradiance. The shape of the curves is most concave around the maximum of $\eta_{Sim,UC}$ at $\kappa = 2.3 \times 10^{20}$ J$^{-1}$ and gets slightly less concave for larger and smaller values of $\kappa$. However, the changes are marginal. At $\kappa = 2.15 \times 10^{20}$ J$^{-1}$ the best agreement between experimental data (black squares) and the simulations was achieved.

## C. Concluding comparison with experiments

In the previous analysis, some input parameters of the upconversion rate equation model have been adapted to experimental data, the absolute value of $\eta_{UC}$ and the fraction between the various

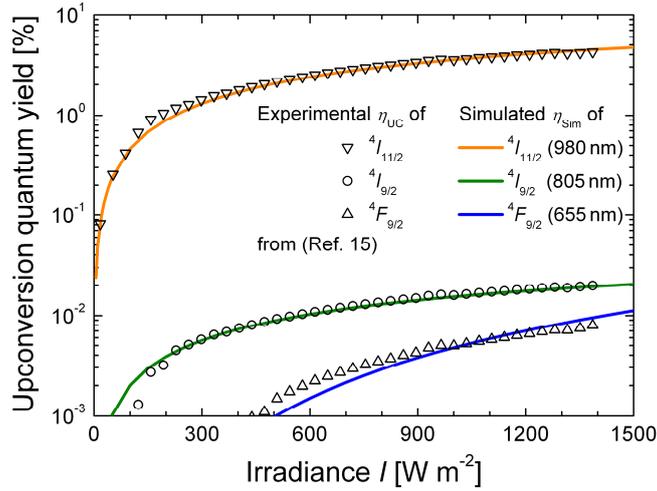

**Figure 13:** Comparison of the simulated internal upconversion quantum yield $\eta_{Sim}$ of various transitions (solid lines) and the corresponding experimentally determined quantum yield $\eta_{UC}$ (open symbols). With the final parameters, optimized at an irradiance of 1000 W m$^{-2}$, the simulations are in good agreement with the experimentally determined $\eta_{UC}$ of the individual transitions from $^4I_{11/2}$, $^4I_{9/2}$ and $^4F_{9/2}$ to the ground state $^4I_{15/2}$

transitions on the $\eta_{UC}$, at an irradiance of 1000 W m$^{-2}$. Additionally, the irradiance dependence of these input parameters was discussed. Figure 13 shows the irradiance dependence for the final input parameters on a logarithmic scale to visualize the contribution of higher energy levels. In contrast to the previous analyses, we compare the irradiance dependence of the individual simulated $\eta_{Sim}$ with the experimentally determined $\eta_{UC}$ of the transitions from $^4I_{11/2}$, $^4I_{9/2}$ and $^4F_{9/2}$ to the ground state $^4I_{15/2}$. The rate equation model with the final parameters describes very well the dependence of the individual transitions on the irradiance.

Many assumptions and estimations are involved to determine the input parameters of the rate equation model, especially for the Einstein coefficients. The analyses, however, show that the nature of upconversion processes can be reasonably described and the comparison in Figure 13 demonstrates that the intensity dependence of the upconverter $\beta$-NaY$_{0.8}$Er$_{0.2}$F$_4$ is well reproduced by our model. Therefore, we aim to use the model for certain applications. For example, a combination of metal nanoparticles with an upconverter is an interesting field of research. With surface plasmons of metal nanoparticles, the UC can be enhanced significantly, depending on the spatial orientation to the metal nanoparticles. Simulations of these nanoparticles can be coupled to our rate equation model to evaluate the possible benefit of such combined materials.

## V.   CONCLUSION

We presented a rate equation model and its processes to describe upconversion of near-infrared photons. The model considers stimulated and spontaneous processes, multi-phonon relaxation and energy transfer of neighboring ions. The parameterization of the model constitutes the main challenge for a physically reasonable simulation of upconversion.

The input parameters for the model were experimentally determined and discussed in detail. With the theory of Kubelka and Munk, the absorption coefficient from powder samples of $\beta$-NaY$_{0.8}$Er$_{0.2}$F$_4$ was calculated. With the Judd-Ofelt analysis, the Einstein coefficients for SPE were determined from the absorption coefficient. The transition probabilities, or Einstein coefficients, are very important parameters to describe the dynamic of the upconversion. For other UC materials or host materials of the Er$^{3+}$ the parameter need to be adjusted, but the fundamental model does not change.

The different processes and parameters within our model influence each other, which makes the analysis complex and challenging. Nevertheless, we showed and explained how the parameters of multi-phonon relaxation and energy transfer affect the $\eta_{Sim}$ of the individual transitions, the contribution of the several $\eta_{Sim}$ on the simulated UC quantum yield $\eta_{Sim,UC}$ and the shape of the irradiance dependence curves of $\eta_{Sim,UC}$. As the results of the simulations with the final parameter set matches the experimental data very well, we conclude that our model reproduces the nature of the upconverter $\beta$-NaY$_{0.8}$Er$_{0.2}$F$_4$. Therefore, we aim to use the model for certain applications, such as evaluate the influence of plasmon resonances of metal-nanoparticles to increase the UC quantum yield and to optimize properties of the host crystal for high UC quantum yields at rather low irradiance.

A broader range of the solar spectrum needs to be utilized by the upconverter to obtain a sufficiently high enhancement of the solar cell efficiency. Thus, UC materials on their own are not very promising to enhance the efficiency of solar cells significantly due to the narrow absorption range attributed to the distinct energy levels. With spectral concentration or downshifting of NIR photons to the $^4I_{15/2}$ to $^4I_{13/2}$ transition energy the effective absorption range of the upconverter can be increased. Luminescent nanocrystalline quantum dots like PbS or core-shell PbSe/PbS have suitable properties to serve as NIR photon downshifting material [64,65].

An irradiance of approximately 140 W m$^{-2}$ of the solar spectrum reaches the earth surface in the spectral range from the band-gap of silicon to a wavelength of 1550 nm, which is suitable for upconversion by Er$^{3+}$. Hence, UC photovoltaic devices with spectral and geometrical concentration of the sunlight [21] are very promising concepts for high efficient solar cells for low concentrated photovoltaic application.

## VI. ACKNOWLEDGEMENT

The authors would like to thank Dr. K. W. Krämer from the University of Bern for helpful discussions and providing the upconverter samples. The research leading to these results has received funding from the German Federal Ministry of Education and Research in the project "Nanovolt - Optische Nanostrukturen für die Photovoltaik" (BMBF, project number 03SF0322H), and from the European Community's Seventh Framework Programme (FP7/2007-2013) under grant agreement n° [246200]. Stefan Fischer gratefully acknowledges the scholarship support from the Deutsche Bundesstiftung Umwelt DBU. Philipp Löper gratefully acknowledges the scholarship support from the Reiner Lemoine Stiftung.